# Spectral Statistics in the Quantized Cardioid Billiard*

by

A. Bäcker† and F. Steiner

II. Institut für Theoretische Physik, Universität Hamburg
Luruper Chaussee 149, 22761 Hamburg
Federal Republic of Germany

and

P. Stifter‡

Abteilung für Quantenphysik, Universität Ulm
89069 Ulm
Federal Republic of Germany

**Abstract:**
The spectral statistics in the strongly chaotic cardioid billiard are studied. The analysis is based on the first 11000 quantal energy levels for odd and even symmetry respectively. It is found that the level-spacing distribution is in good agreement with the GOE distribution of random-matrix theory. In case of the number variance and rigidity we observe agreement with the random-matrix model for short-range correlations only, whereas for long-range correlations both statistics saturate in agreement with semiclassical expectations. Furthermore the conjecture that for classically chaotic systems the normalized mode fluctuations have a universal Gaussian distribution with unit variance is tested and found to be in very good agreement for both symmetry classes. By means of the Gutzwiller trace formula the trace of the cosine-modulated heat kernel is studied. Since the billiard boundary is focusing there are conjugate points giving rise to zeros at the locations of the periodic orbits instead of exclusively Gaussian peaks.

*Supported by Deutsche Forschungsgemeinschaft under Contract No. DFG-Ste 241/7–1.
†E-mail address: `baeckera@x4u2.desy.de`
‡E-mail address: `stif@physik.uni-ulm.de`

chao-dyn/9412007  14 Dec 1994

# 1 Introduction

Today the important role played by chaos in nonlinear dynamical systems is generally appreciated. The most striking property of deterministic chaos is the sensitive dependence on initial conditions such that neighbouring trajectories in phase space separate at an exponential rate. As a result the long-time behaviour of a strongly chaotic system is unpredictable. There the fundamental question arises whether this well-established phenomenon of classical chaos has an analogue in the quantum world that could be called "quantum chaos". By this the following is meant: Given a dynamical system which is strongly chaotic, i.e., ergodic, mixing and a K-system, is there any manifestation in the corresponding quantal system which betrays its chaotic character? (For an authoritative review, see ref. [1].) If one were to identify unique fingerprints of classical chaos in quantum mechanics, one could use these to *define* quantum chaos. Ideally, classically chaotic systems should be characterized by a random behaviour of these fingerprints that qualify the system to be called "chaotic" also in quantum mechanics.

For bound conservative systems, the quantum mechanical time evolution is almost periodic, in the sense of Harald Bohr's theory of almost periodic functions, due to the discrete spectrum of the time-evolution operator. One thus observes no sensitive dependence on initial conditions in the long-time behaviour of quantum mechanics. This is in contrast to classical systems whose time evolution is ruled by the Liouville operator. If the classical dynamics is strongly chaotic, the spectrum of the Liouvillian has a continuous spectrum on the unit circle which leads to a decay of time correlations of classical observables reflecting a complete loss of information on the system. This fundamental difference between classical and quantum mechanics has led to the common belief that there hardly exists any phenomenon in quantum mechanics which justifies the notion of quantum chaos.

However, instead of concentrating on the long-time behaviour, one can consider the extreme limit $t = \infty$ in quantum mechanics and thus study properties of stationary states, that is of eigenvalues and eigenfunctions of the corresponding time-independent Hamiltonian. The idea is that the statistical properties of the energy-level fluctuations ("spectral statistics") and wave functions of a given quantum system are already determined by its classical limit, depending only upon whether this limit is chaotic or not. (In this paper we shall consider the spectral statistics only; for a discussion of the statistical properties of wave functions in chaotic systems, see refs. [1, 2, 3, 4, 5, 6].) It has been conjectured that the statistical properties of quantum energy spectra of classically integrable systems can be described [7] by Poissonian random processes, whereas the spectral statistics of strongly chaotic systems can be described [8] by the universal laws of random-matrix theory (RMT), originally proposed by Wigner, and Landau and Smorodinsky, and fully developed by Dyson for a better understanding of the resonances of compound nuclei. (See ref. [9] for a collection of the original papers, and refs. [1, 10, 11] for recent reviews.) The *random-matrix model* for spectral statistics has to be viewed as a purely phenomenological one, in contrast to random-matrix theory which constitutes an exact mathematical theory. So far there does not exist a complete theory for the spectral statistics of chaotic systems, nor even for integrable ones. Based on Gutzwiller's trace formula [12, 1], Berry carried out a semiclassical analysis [13, 14, 15] of the level fluctuations and obtained a saturation of the two-point statistics for long-range correlations in contradiction with the predictions based on RMT; non-universal long-range correlations occur due to the non-universal behaviour of short periodic orbits. A recent analysis [16, 17] based on the exact Selberg trace formula has clearly confirmed the non-universal saturation up to correlation lengths as large as $L = 700$ (in units of the average level spacing). In addition, there exists a class of strongly chaotic systems for which the spectral fluctuations nearly behave as it is expected for classically integrable systems. This phenomenon occurs for geodesic flows on hyperbolic manifolds (for instance, Riemannian surfaces with constant negative Gaussian curvature), whose fundamental groups are of an arithmetical origin; thus the notion of arithmetical chaos was introduced [18, 19]. The dynamical systems possessing arithmetical chaos violate universality in energy-level statistics even in the short-range regime. It thus appears that the random-matrix model does not provide a universal signature of classical chaos in quantum mechanics.

Recently, a novel quantity to measure quantum chaos has been proposed and a conjecture about its statistical behaviour has been put forward [5, 6]. According to this conjecture there are unique fluctuation properties in quantum mechanics which are *universal* and, in a well-defined sense, *maximally random* if the corresponding classical system is strongly chaotic. Numerical as well as theoretical evidence has been provided in favour of the conjecture [6, 17]. A rigorous proof of the conjecture would give us a clear-cut definition of *quantum chaos in spectra*.

In order to shed more light on the relationship between classically chaotic systems and the corresponding quantum systems it is important to have a large number of systems for which it is possible to carry out extensive numerical computations of the relevant spectral statistics. The simplest nonlinear dynamical systems one can study are billiard systems. They consist of a point particle moving freely inside a curved boundary with elastic reflections at the boundary. These systems are specially suited for the study of a possible manifestation of classical chaos in quantum mechanics, because a lot of mathematical results are available. Furthermore the integration of the equations of motion is trivial, i.e., the geodesics are straight lines, which by no means does imply that it is trivial to determine all periodic orbits having periods below a given value. Moreover billiard systems belong to the so-called scaling systems with the semiclassical limit of Planck's constant $\hbar \to 0$ being equivalent to the high energy limit $E \to \infty$.

In this paper we are concerned with a billiard system bounded by the cardioid [20]. This system is of special interest because it provides one of the few examples for which it is rigorously proven that it is strongly chaotic [21, 22, 23, 24]. Famous other examples are the Sinai billiard [25, 26], which is very important for the foundations of statistical mechanics, and Bunimovich's stadium billiard [27, 28, 29, 30]. However, the last two systems possess some non-generic features, because they have a family of stable periodic orbits, the so-called bouncing-ball modes. The cardioid billiard has no such family, and no whispering gallery orbits exist. Furthermore the boundary is focusing and the geometric properties are such that caustics exist. Their important influence will become clear in sec. 4 in the discussion of the cosine-modulated heat kernel.

The paper is organized as follows. After defining the cardioid billiard and its quantum mechanical version in sec. 2, we present in sections 3.1 to 3.5 a detailed analysis of the spectral statistics, i.e., the spectral staircase function, the $\delta_n$-statistic, the level-spacing distribution $P(s)$, the number variance $\Sigma^2(L)$ and the rigidity $\Delta_3(L)$. In sec. 3.6 the distribution of the mode-fluctuation number is calculated and found to be in excellent agreement with the recent conjecture [5, 6] on quantum chaos in spectra. In order to keep the number of figures limited, the various spectral measures are displayed in each case for one symmetry class only, but we show the plots for even and odd symmetry in an alternating order. In section 4 we consider the trace of the cosine-modulated heat kernel as a an additional test of the accuracy of the eigenvalues and as a first application of Gutzwiller's periodic-orbit theory. Section 5 contains a summary and discussion of our results.

# 2 The Cardioid Billiard

The billiard system to be studied is given by the free motion of a point particle inside a two-dimensional Euclidean domain $\Omega$ bounded by the cardioid (see figure 1) with elastic reflections at the boundary $\partial\Omega$. The cardioid billiard is the limit case of a family of billiards which was introduced by Robnik [20]. The boundary $\partial\Omega$ of the billiards is defined by a quadratic conformal mapping of the unit circle

$$w = u + iv = z + \lambda z^2 \quad ; \quad z = e^{i\varphi} \, , \quad \varphi \in [0, 2\pi[ \, , \quad \lambda \in [0, \tfrac{1}{2}[ \, . \tag{1}$$

Starting with $\lambda = 0$ one gets a continuous deformation of the circle. The cardioid billiard is obtained for $\lambda = \frac{1}{2}$; for this value the mapping is no longer conformal, since $\frac{dw}{dz} = 1 + z = 0$ for $z = -1$ ($\varphi = \pi$), where the cardioid has a cusp. It was proven that the cardioid billiard is ergodic, mixing, and a K-system [21, 22, 23]. In fact it is even a Bernoulli system, which follows from a recent result [24].

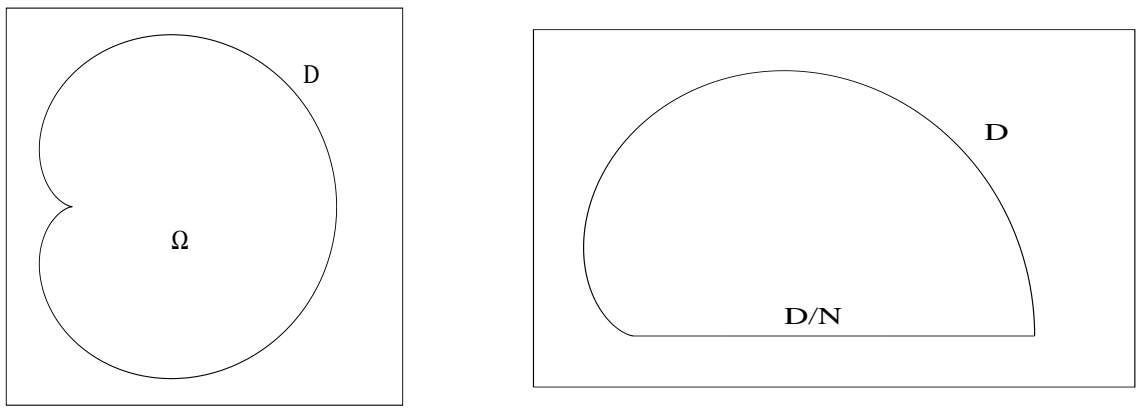

Figure 1: Full and desymmetrized cardioid billiard. D denotes Dirichlet and N Neumann boundary conditions.

The corresponding quantum mechanical system is governed by the Schrödinger equation (in natural units $\hbar = 2m = 1$)

$$-\Delta \Psi_n(\vec{q}) = E_n \Psi_n(\vec{q}) \quad , \quad \vec{q} \in \Omega \, ,$$

with Dirichlet condition $\Psi_n(\vec{q})=0$ at the boundary $\partial\Omega$. $\Delta$ is the two-dimensional Laplacian. Furthermore, we have the orthonormality relation

$$\int_\Omega d^2q \, \Psi_m(\vec{q}) \, \Psi_n(\vec{q}) = \delta_{mn} \, .$$

Notice that the eigenfunctions $\Psi_n(\vec{q})$ can be chosen as real. For $\Omega$ being compact the energy spectrum $\{E_n\}$ is purely discrete. Since the system is invariant under reflection at the symmetry line ($v = 0$) one can classify the wave functions $\Psi_n(\vec{q})$ as odd and even eigenfunctions satisfying Dirichlet or Neumann boundary conditions respectively at the symmetry line. Therefore we will consider the two desymmetrized versions of the cardioid billiard only, see figure 1.

For the following discussion it is important to keep in mind that the semiclassical limit $\hbar \to 0$ corresponds to the high energy limit $E_n \to \infty$.

# 3 Spectral Statistics

## 3.1 Spectral Staircase and Weyl's Law

One can easily check the asymptotic behaviour of the spectrum by calculating the spectral staircase function (integrated level density)

$$N(E) = \#\{n | E_n < E\} \tag{2}$$

which counts the number of energy levels below $E$. $N(E)$ can be divided into a smooth and an oscillatory part

$$N(E) = \bar{N}(E) + N_{\text{osc}}(E) \ .$$

The mean behaviour of $N(E)$ is asymptotically for $E \to \infty$ described by the generalized Weyl's law including perimeter, corner and curvature corrections [31]. For odd $(-)$ and even $(+)$ symmetry respectively, one obtains

$$\bar{N}^-(E) = \frac{3}{16}E - \frac{3}{2\pi}\sqrt{E} + \frac{3}{16} \ , \tag{3}$$

$$\bar{N}^+(E) = \frac{3}{16}E - \frac{1}{2\pi}\sqrt{E} \ , \tag{4}$$

where the inward pointing cusp contributes like an edge with angle $\pi$ in the desymmetrized billiard.

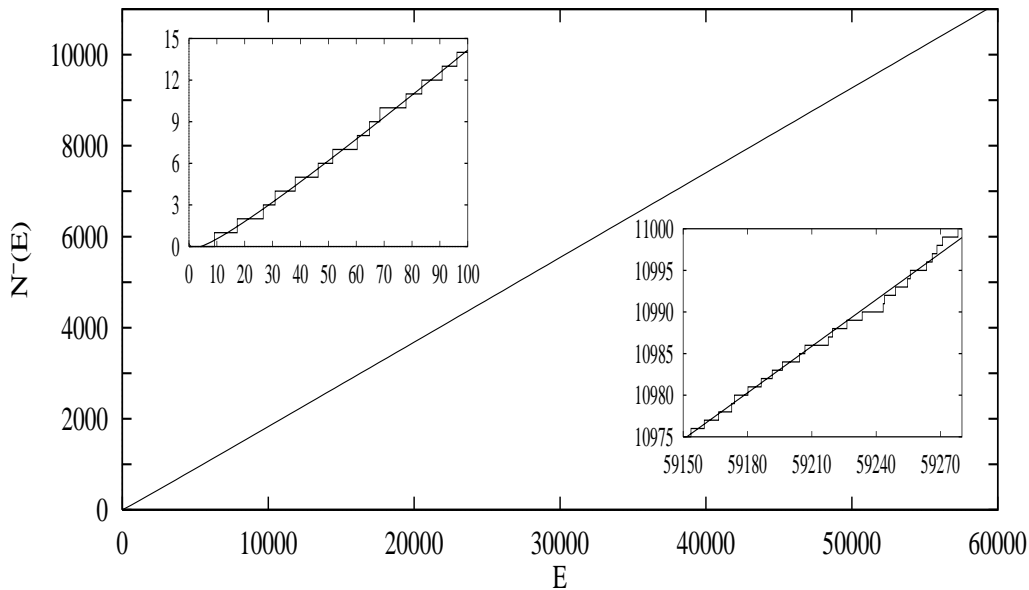

Figure 2: Spectral staircase $N^-(E)$ for odd symmetry and generalized Weyl's law, equation (3). The insets show a magnification of the energy intervals [0,100] and [59150,59270] respectively.

In figure 2 the spectral staircase $N^-(E)$ is shown for the first 11000 energy levels for odd symmetry and compared with the generalized Weyl's law $\bar{N}^-(E)$, equation (3). Due to the large energy interval no difference between both curves is visible in this figure on the whole range. Therefore two insets at the lowest and highest end are shown. The eigenvalues were computed by Prosen and Robnik [32] by using the conformal mapping diagonalization technique which was introduced by Robnik [33, 34]. In ref. [34] the energy-level statistics were investigated in detail for the family of billiards (1) for various parameters $\lambda < \frac{1}{2}$ for which a rigorous proof of strong chaos is lacking.

Despite the fact that $\bar{N}^-(E)$ is an asymptotic law, it gives the correct mean behaviour down to the ground state, as it was observed in many other systems before. Figure 2 strongly indicates that the computed energy spectrum for odd symmetry is complete for the first 11000 levels. In the same way we have also checked the energy spectrum for even symmetry being complete for the first 11000 levels. Furthermore no degeneracies of eigenvalues in the spectra for even and odd symmetry were found.

In order to compare the eigenvalues of the cardioid billiard with those of other systems, it is necessary to unfold the spectrum by means of the generalized Weyl's law (3) or (4)

$$E_n'^{\pm} = \bar{N}^{\pm}(E_n^{\pm}) \; .$$

In the following we shall analyze the unfolded spectra $\{E_n'^{\pm}\}$ for the two desymmetrized billiards, but the prime and superscript $\pm$ will be omitted. After the process of unfolding the spectrum has a mean level spacing of unity, and different systems differ in the oscillating part $N_{\rm osc}(E)$ only.

## 3.2 $\delta_n$-Statistic

A more refined way of testing whether the spectrum is complete can be performed by considering the fluctuating part $N_{\rm osc}(E_n)$ of the spectral staircase function evaluated[1] at the unfolded eigenvalues $E_n$

$$\delta_n := N_{\rm osc}(E_n) := N(E_n) - \bar{N}(E_n) = n - \tfrac{1}{2} - E_n \; . \qquad (5)$$

As one can see in figure 3, where the even case is shown, $\delta_n$ is oscillating around zero, showing that the levels indeed obey the mean behaviour as described by Weyl's law. If a level was missing, $\delta_n$ would oscillate around $-1$ from then on. This indicates that the spectrum for even symmetry is complete for the first 11000 levels. The same holds for the odd case.

Remarkably, the $\delta_n$-statistic is even sensitive to the constant term in Weyl's law, equations (3) and (4), as one can infer from the mean values of $\delta_n$ for which we obtain $\langle\delta\rangle = -0.00072$ for odd, and $\langle\delta\rangle = -0.00217$ for even symmetry. This is in the odd case more than a factor 200 less than the constant $\mathcal{C}_{\rm odd} = \frac{3}{16} = 0.1875$ in equation (3). In case of the constant being neglected or unknown, the mean value of $\delta_n$ would provide a good approximation to the constant term in Weyl's law.

Although the fluctuations $\delta_n$ have mean zero, a look at figure 3 suggests that their variance slowly increases with increasing energy. A quantitative discussion of this important observation will be postponed until section 3.5.

---

[1] Here the value of the spectral staircase function (2) at an unfolded eigenvalue $E_n$ is defined as

$$N(E_n) = \lim_{\epsilon \to 0} \left( \frac{N(E_n - \epsilon) + N(E_n + \epsilon)}{2} \right) = n - \tfrac{1}{2} \; .$$

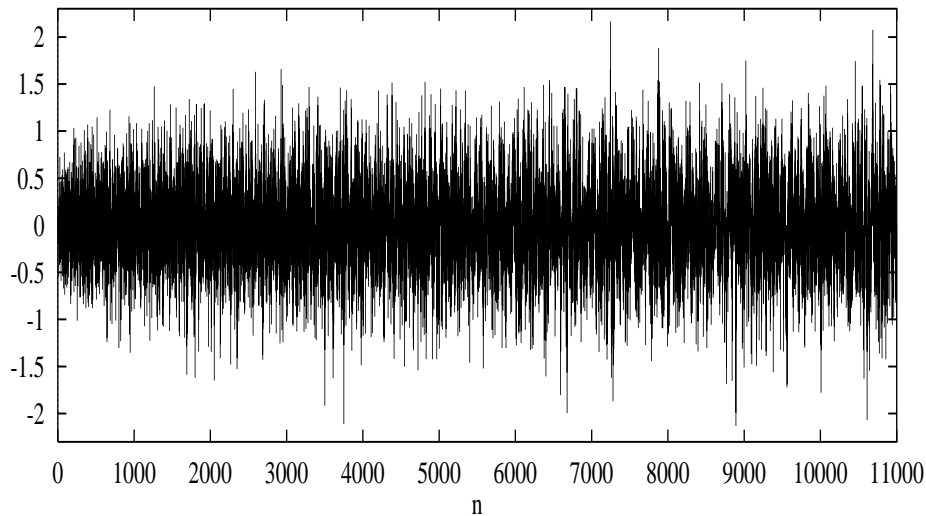

Figure 3: Plot of the fluctuating part $\delta_n := N_{\text{osc}}(E_n) := n - \frac{1}{2} - E_n$ for even symmetry for the first 11000 eigenvalues

## 3.3 Level-Spacing Distribution

An important statistic measuring short-range correlations of the spectrum is the nearest-neighbour level-spacing statistic $P(s)$. $P(s)\,ds$ is the probability of finding an arbitrary pair of nearest neighbours of energy levels with spacing $s_n = E_{n+1} - E_n$ in the interval $[s, s + ds]$. Thus $P(s)$ measures fluctuations of the distances between two nearest-neighbour levels.

The level-spacing distribution for typical integrable systems is found to obey the Poissonian distribution $P^{\text{POIS}}(s) = e^{-s}$, which is in clear contrast to the level repulsion observed in generic chaotic systems. The random-matrix model [10] predicts that for systems with time-reversal symmetry the level-spacing distribution $P(s)$ is given by the distribution of the Gaussian orthogonal ensemble (GOE) which is well described by a Wigner distribution

$$P^{\text{GOE}}(s) = \frac{\pi}{2} s \exp\left\{-\frac{\pi}{4} s^2\right\} \quad ,$$

whereas systems without such symmetry should obey the distribution of the Gaussian unitary ensemble (GUE) which is in good approximation given by

$$P^{\text{GUE}}(s) = \frac{32}{\pi^2} s^2 \exp\left\{-\frac{4}{\pi} s^2\right\} \quad .$$

Since the cardioid billiard possesses time-reversal symmetry the level-spacing distribution is expected to be described by a Wigner distribution.

In figure 4 the level-spacing distribution for odd symmetry is shown. The agreement with $P^{\text{GOE}}(s)$ is very good. A similar result holds in the even symmetry case.

The level-spacing distribution has the drawback that there is a loss of information due to the binning. Therefore it is more significant to look at the cumulative level-spacing distribution

$$I(s) = \int_0^s P(s')\,ds' \quad .$$

Numerically this quantity is calculated by sorting the set of nearest-neighbour spacings $\{s_n\}$ and determining the fraction of level spacings below a given $s$.

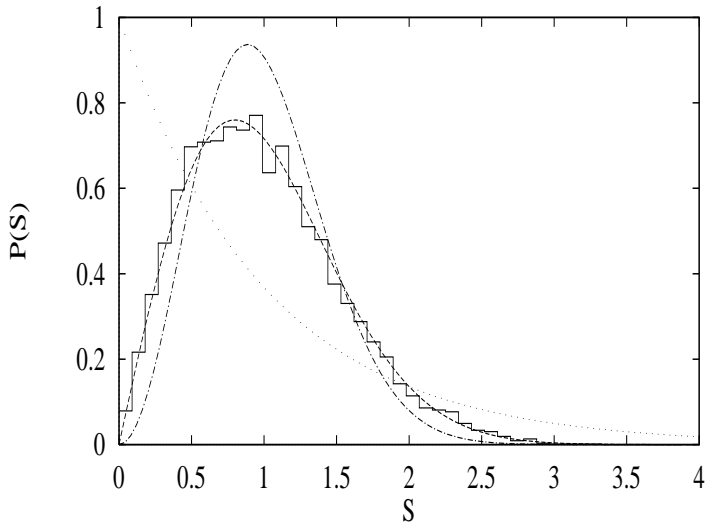

Figure 4: Histogram of the nearest-neighbour level-spacing distribution $P(s)$ for odd symmetry. The dashed line is the expected Wigner distribution, the dashed-dotted line is the GUE distribution, and the dotted line shows the Poissonian distribution expected for integrable systems.

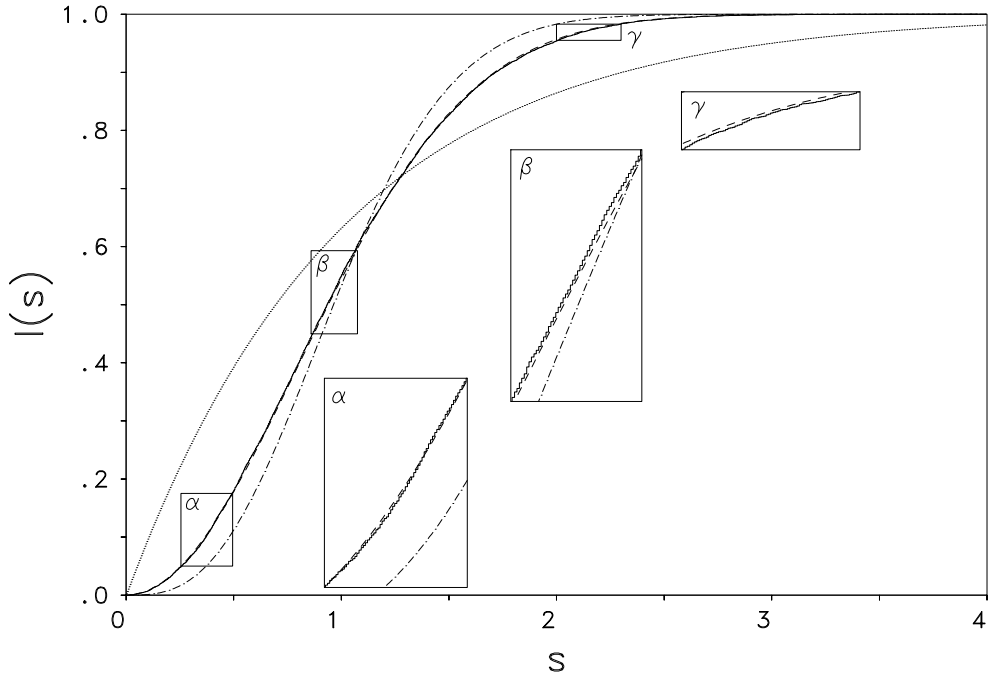

Figure 5: Cumulative level spacing for even symmetry. Full line: calculated from the energy levels, dashed line: GOE expectation $I^{\mathrm{GOE}}(s)$, dashed-dotted line: GUE expectation $I^{\mathrm{GUE}}(s)$, and dotted line the Poissonian distribution $I^{POIS}(s)$.

As one can see in figure 5, where the even case is shown, the agreement between the cumulative level-spacing distribution and the cumulative GOE expectation is very good. The Kolmogorov-Smirnov test gives a significance level of $\mathcal{P} = 24\%$ for the odd, and $\mathcal{P} = 67\%$ for the even case assuming a GOE distribution, whereas for the GUE distribution an extremely small significance level of $\mathcal{P} < 10^{-52}$ is obtained for both symmetry classes.

## 3.4 Number Variance

In contrast to the level-spacing statistic, the number variance allows one to investigate also medium- and long-range correlations of the spectrum. It is defined by

$$\Sigma^2(L, \hat{E}) := \left\langle (n(L, E) - L)^2 \right\rangle_{\hat{E}, \delta} \qquad , \qquad L > 0 \quad ,$$

which is the local variance of the number $n(L, E) = N(E + \frac{L}{2}) - N(E - \frac{L}{2})$ of unfolded energy levels $E_n$ in the interval $[E - \frac{L}{2}, E + \frac{L}{2}]$. The brackets $\langle \ldots \rangle_{\hat{E}, \delta}$ denote a local average with center $\hat{E}$ and effective width $\delta$. Different averaging procedures are possible; here we have chosen a rectangular averaging

$$\langle f(E) \rangle_{\hat{E}, \delta} = \frac{1}{\delta} \int_{\hat{E} - \delta/2}^{\hat{E} + \delta/2} f(E) \, dE \quad . \tag{6}$$

In random-matrix theory one obtains in the GOE case

$$\begin{aligned}\Sigma^2_{\text{GOE}}(L) &= \frac{2}{\pi^2} \left\{ \log(2\pi L) + \gamma + 1 + \frac{1}{2}\text{Si}^2(\pi L) - \frac{\pi}{2}\text{Si}(\pi L) \right. \\ &\quad \left. - \cos(2\pi L) - \text{Ci}(2\pi L) + \pi^2 L \left[ 1 - \frac{2}{\pi}\text{Si}(2\pi L) \right] \right\} \quad ,\end{aligned} \tag{7}$$

and for GUE

$$\Sigma^2_{\text{GUE}}(L) = \frac{1}{\pi^2} \left\{ \log(2\pi L) + \gamma + 1 - \cos(2\pi L) - \text{Ci}(2\pi L) + \pi^2 L \left[ 1 - \frac{2}{\pi}\text{Si}(2\pi L) \right] \right\} \quad , \tag{8}$$

where $\text{Si}(x)$ and $\text{Ci}(x)$ are the sine and cosine integral respectively, and $\gamma = 0.5772\ldots$ is Euler's constant. For a random Poissonian process one has

$$\Sigma^2_{\text{POIS}}(L) = L \quad ,$$

which is in agreement with the general small-$L$ behaviour $\Sigma^2(L) = L + O(L^2)$, following from the fact that $N(E)$ is a staircase function.

In figure 6 we show the number variance in the odd symmetry case for $L \leq 50$ and different values of $\hat{E}$: $\hat{E} = 1000, 4000, 7000, 10000$ ($\delta = 1800$). In table 1 the saturation values $\Sigma^2_\infty(\hat{E})$, calculated as the average of $\Sigma^2(L, \hat{E})$ over $L \in [15, 100]$, are listed. It is seen that the agreement with the GOE expectation (dashed-dotted line) is restricted to $L \leq 2 \ldots 7$, depending on $\hat{E}$. Analogous results hold in the even symmetry case.

It is now well established that the medium- and long-range statistics measured by the number variance strongly depend on the (non-universal) short periodic orbits [13, 14, 15, 35, 16, 17]. Therefore the universality regime, which is expected to be governed by the random-matrix model, is restricted to very small correlation lengths $L$, which is confirmed by our numerical results. In order to enlarge the range of agreement with the random-matrix model, the center of the averaging window has to be in the deep semiclassical regime corresponding to extremely high lying energy levels.

According to Berry's semiclassical analysis [14, 15] $\Sigma^2_\infty(\hat{E})$ should increase with increasing energy $\hat{E}$ in the limit of large $\hat{E}$. A look at figure 6 and table 1 shows that such an overall increase is indeed observed within the relatively large fluctuations of the number variance.

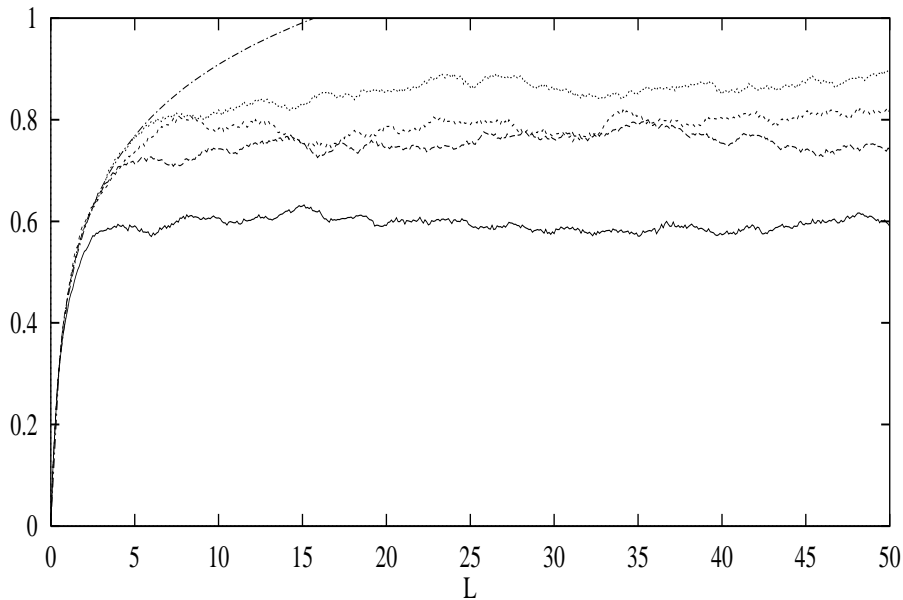

Figure 6: Number variance $\Sigma^2(L, \hat{E})$ for odd symmetry at different energies $\hat{E}$: $\hat{E} = 1000$ (full line), 4000 (long dashed line), 7000 (short dashed line), 10000 (dots); $\delta = 1800$. Dashed-dotted line: GOE expectation $\Sigma^2_{\text{GOE}}(L)$.

| $\hat{E}$ | $\Sigma^2_\infty(\hat{E})$ (odd) | $\Sigma^2_\infty(\hat{E})$ (even) | $\Delta_\infty(\hat{E})$ (odd) | $\Delta_\infty(\hat{E})$ (even) |
|---|---|---|---|---|
| 1000 | 0.59 | 0.60 | 0.29 | 0.29 |
| 4000 | 0.75 | 0.75 | 0.38 | 0.37 |
| 7000 | 0.79 | 0.83 | 0.40 | 0.41 |
| 10000 | 0.85 | 0.83 | 0.43 | 0.38 |

Table 1: Saturation values $\Sigma^2_\infty(\hat{E})$ of the number variance, and $\Delta_\infty(\hat{E})$ of the rigidity at different energies $\hat{E}$. Notice that the last digit is uncertain.

## 3.5 Spectral Rigidity

Another statistic measuring two-point correlations is the spectral rigidity [10]

$$\Delta_3(L, \hat{E}) := \left\langle \min_{(a,b)} \frac{1}{L} \int_{-L/2}^{L/2} d\epsilon \, [N(E + \epsilon) - a - b\epsilon]^2 \right\rangle_{\hat{E}, \delta} , \quad (9)$$

which is the average mean square deviation of the spectral staircase function from the best fitting straight line over an energy range of $L$ mean level spacings. The brackets $\langle \ldots \rangle_{\hat{E}, \delta}$ again represent a local average.

For short correlation lengths $L \ll 1$ the rigidity is independent of the underlying spectrum, and obeys $\Delta_3(L) \sim \frac{1}{15}L$, which is a consequence of $N(E)$ being a staircase function.

In the theory of random matrices (RMT) $\Sigma^2(L)$ and $\Delta_3(L)$ are related through

$$\Delta_3^{\text{RMT}}(L) = \frac{2}{L^4} \int_0^L \left( L^3 - 2L^2 r + r^3 \right) \Sigma^2_{\text{RMT}}(r) \, dr . \quad (10)$$

By this equation it is possible to obtain $\Delta_3^{\mathrm{GOE}}(L)$ or $\Delta_3^{\mathrm{GUE}}(L)$ by numerical integration of equations (7) or (8).

In figure 7 the rigidity calculated from equation (9) is shown for different energies $\hat{E}$ and $\delta = 1800$. One clearly observes a saturation effect of the rigidity for large $L$ as it is suggested by Berry's analysis [13] of the spectral rigidity using the semiclassical trace formula. This is in contrast to the logarithmic increase predicted by random-matrix theory, see equation (7). The rigidity is smoother than the number variance, but reaches the saturation later in comparison with $\Sigma^2(L, \hat{E})$.

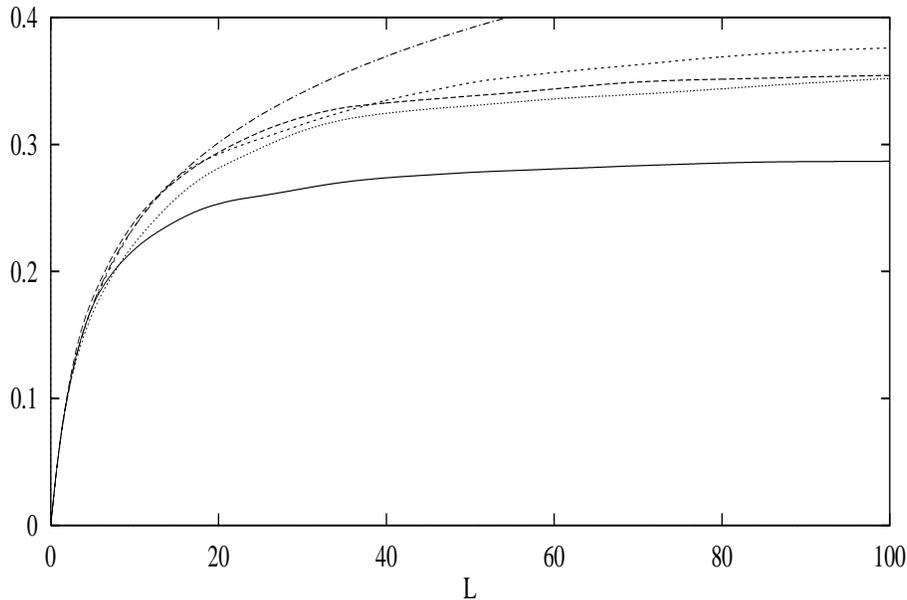

Figure 7: Rigidity $\Delta_3(L, \hat{E})$ for even symmetry at different energies $\hat{E}$: $\hat{E} = 1000$ (full line), 4000 (long dashed), 7000 (short dashed), 10000 (dots), $\delta = 1800$. The dashed-dotted line shows the expectation of random-matrix theory in the GOE case $\Delta_3^{\mathrm{GOE}}(L)$.

The prediction of the random-matrix model for the rigidity suffers from the same drawback as in case of the number variance, i.e., that it is valid for small correlation lengths $L$ only.

For the following section it is of great importance to know the dependence of the saturation value $\Delta_\infty(\hat{E})$ of the rigidity on the energy. $\Delta_\infty(\hat{E})$ is defined as $\Delta_\infty(\hat{E}) = \lim_{L \to \infty} \Delta_3(L, \hat{E})$. According to Berry's semiclassical analysis [13] one expects in the semiclassical limit $\hat{E} \to \infty$ for systems with time-reversal symmetry

$$\Delta_\infty(\hat{E}) = \frac{1}{2\pi^2} \log \hat{E} + C \quad . \tag{11}$$

The constant $C$ has been estimated by Berry as $C = C(l_0) = \frac{1}{\pi^2} \log(4\pi e \bar{d}(\hat{E})/l_0) - \frac{1}{8}$ where $\bar{d}(\hat{E}) = \frac{d\bar{N}(\hat{E})}{d\hat{E}}$ is the mean level density, $l_0$ denotes the length of the shortest periodic orbit and $e$ is the base of the natural logarithm. In case of the cardioid billiard we have $l_0 = 2.598\ldots$ and thus obtain $C(l_0) \approx -0.034$ for both symmetry classes. Due to the finite number of available energy levels we cannot perform the limit $L \to \infty$. Therefore we have determined $\Delta_\infty(\hat{E})$ from a fit of $\Delta_3(L, \hat{E})$ in the range $L \in [15, 400]$ for fixed $\hat{E}$ to the function

$$\Delta_3^{fit}(L) = \Delta_\infty \left(1 + \frac{a}{L} + \frac{b}{L^2}\right) \quad , \tag{12}$$

where $\Delta_\infty$, $a$ and $b$ are fit parameters. We have chosen 9 different energies $\hat{E}$ from the interval $[1000, 10000]$ and determined the corresponding saturation values $\Delta_\infty(\hat{E})$ using the fit (12). Fitting in turn the asymptotic behaviour (11) to these values we obtained for the constant $C = -0.048$ in the odd, and $C = -0.052$ in the even case. The result of the saturation values together with the asymptotic curve (11) are represented for the odd case in figure 8. Notice that the values obtained for the constant $C$ deviate from the simple estimate using only the shortest periodic orbit.

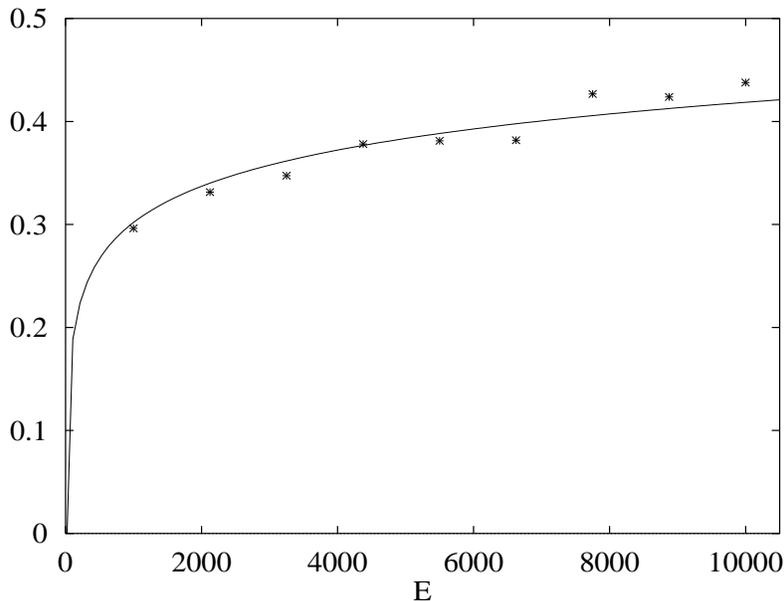

Figure 8: Saturation values $\Delta_\infty(E)$ (stars) for odd symmetry at different energies, and the asymptotic curve (11) with $C = -0.048$ (full line).

Another interesting aspect which can be deduced from equation (10) is that there exists a simple relation between the two saturation values [35]

$$\Sigma^2_\infty = 2\Delta_\infty \ . \tag{13}$$

This relation is confirmed by our numerical data within the expected errors, see table 1, except at $\hat{E} = 10000$ in the even case.

## 3.6 The Mode-Fluctuation Distribution $P(W)$

As we have seen in the preceding sections, the agreement of the computed number variance and rigidity with the predictions of random-matrix theory holds only at small correlation lengths. A novel quantity which can be used as an indicator of quantum chaos in spectra is the distribution $P(W)$ of the mode fluctuation $W(E)$ [5, 6]. The function $W(E)$ is defined as the normalized fluctuations of the mode number $N(E)$ around the mean mode number $\bar{N}(E)$

$$W(E) := \frac{N(E) - \bar{N}(E)}{\sqrt{\Delta_\infty(E)}} = \frac{N_{\text{osc}}(E)}{\sqrt{\Delta_\infty(E)}} \ . \tag{14}$$

|  | odd symmetry | even symmetry | conjecture (15) |
|---|---|---|---|
| Average | −0.001 | −0.003 | 0 |
| Variance | 0.992 | 0.996 | 1 |
| Skewness | −0.021 | 0.008 | 0 |
| Kurtosis | 0.112 | −0.089 | 0 |

Table 2: First moments of the mode-fluctuation distribution $P(W)$ for the odd and even case.

By definition $N_{\text{osc}}(E)$ fluctuates around zero, see figure 3, and furthermore it can be shown [19] that the second moment of $N_{\text{osc}}(E)$ tends asymptotically to the saturation value $\Delta_\infty(E)$ of the rigidity.

In [5, 6] the conjecture has been put forward that classically chaotic systems should display a universal Gaussian behavior in the limit $E \to \infty$

$$P_{Gauss}(W) = \frac{1}{\sqrt{2\pi}} e^{-\frac{1}{2}W^2} \quad , \tag{15}$$

whereas classically integrable systems should display non–Gaussian distributions $P(W)$.

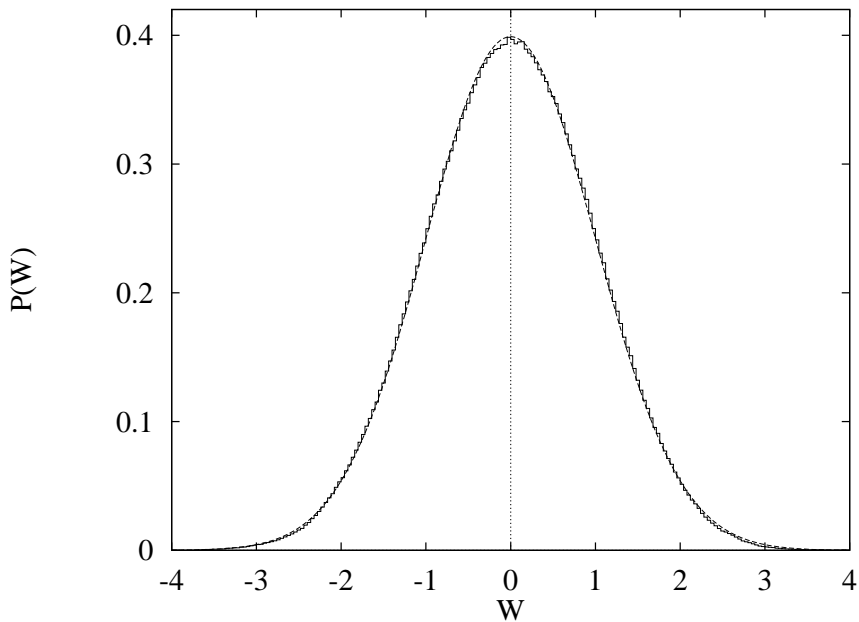

Figure 9: Histogram of the mode-fluctuation distribution $P(W)$ (odd case). Dashed line: Gaussian limit distribution.

Numerically the distribution function was calculated by randomly choosing $10^7$ energy values $E \in [0, 11000]$ and determining the histogram of $W(E)$. As one can see in figure 9 the agreement between the histogram and the Gaussian normal distribution is very good. In table 2 the mean value, variance, skewness and kurtosis for the odd and even cases are shown.

In order to quantify the visual impression, we applied the Kolmogorov–Smirnov test to the cumulative mode-fluctuation distribution $C(W)$ (see figure 10). We have chosen equidistantly 11000 points of the unfolded energy interval $[0, 11000]$ and determined the maximum value $D$ of the absolute difference between the cumulative distribution and the cumulative normal distribution. The significance level $\mathcal{P}$ of the distance $D$ is in both symmetry cases above 60%, which strongly confirms the conjecture.

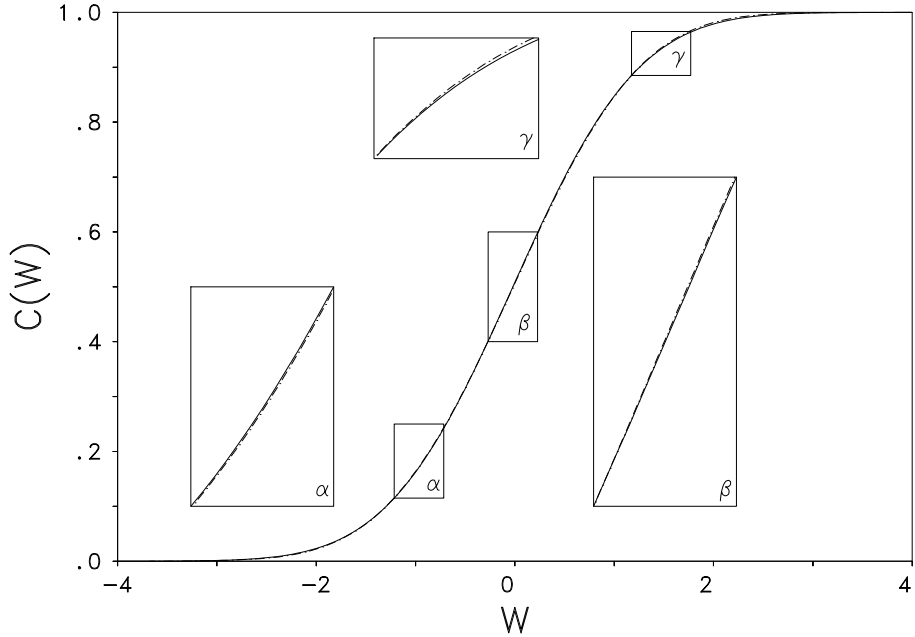

Figure 10: Cumulative mode-fluctuation distribution for the even case in comparison with the cumulative version of (15).

## 4 The Trace of the Cosine–Modulated Heat Kernel

It is well known that the Gutzwiller trace formula for bound quantum systems [12, 1] obtained from a semiclassical approximation for the density of states is at best conditionally convergent. Based on Gutzwillers semiclassical approximation for the level density a generalized trace formula has been derived by considering the trace $\text{Tr}(h(\hat{H}^{1/2}))$ of a suitable test function $h(p)$ of the square root of the Hamiltonian $\hat{H}$ [36]. The result obtained in the semiclassical limit of Planck's constant $\hbar \to 0$ is a periodic-orbit sum rule, which involves absolutely convergent sums and integrals only ($\hbar = 2m = 1$)

$$\sum_{n=1}^{\infty} h(p_n) = \int_0^{\infty} dp\, h(p)\bar{d}(p) + \mathcal{C}\, h(0) + \sum_{\gamma} \sum_{k=1}^{\infty} \frac{l_\gamma}{\sqrt{|2 - \text{Tr}M_\gamma^k|}} \mathcal{F}_\gamma\{h(kl_\gamma)\} \ . \qquad (16)$$

Here the sum on the left-hand side runs over all quantal eigenvalues (not unfolded !) parametrized by the discrete momenta $\{p_n = \sqrt{E_n}\}$. The integral on the right-hand side is the so-called "zero–length contribution" and is completely determined by the generalized Weyl's law $\bar{d}(p) = \frac{d\bar{N}(p^2)}{dp}$. $\mathcal{C}$ is the constant term in Weyl's law.

The $\gamma$-summation runs over all primitive periodic orbits $\gamma$ of length $l_\gamma$, whereas $\sum_k$ counts multiple traversals corresponding to periodic orbits of length $kl_\gamma$; $M_\gamma$ is the monodromy matrix, and $\mathcal{F}_\gamma\{h(x)\}$ denotes the Fourier transform of the test function $h(p)$ (incorporating the phase shift due to the Maslov index)

$$\mathcal{F}_\gamma\{h(x)\} = \frac{1}{\pi} \int_0^\infty dp\, h(p)\, \cos(px - k\nu_\gamma \tfrac{\pi}{2}) \;, \tag{17}$$

where $\nu_\gamma$ denotes the Maslov index which is twice the number of reflections at Dirichlet boundaries plus the maximal number of conjugate points $\mu_\gamma$ (see e.g. [37, 38]).

Starting from the generalized trace formula (16) we can study the inverse problem of quantum chaos: to obtain the lengths of the periodic orbits from the quantal energy levels. This problem can be solved by considering the trace of the cosine–modulated heat kernel [39, 40] $\mathrm{Tr}\{\cos\left((-\Delta)^{1/2} L\right) e^{t\Delta}\}$, which is obtained from equation (16) by choosing the following test function

$$h(p) = \cos(pL)\, e^{-p^2 t}\,; \qquad p = \sqrt{E}\,,\quad t > 0 \;. \tag{18}$$

Due to the geometry of the cardioid billiard with its focusing boundary, the orbits listed in table 3 have a non–vanishing maximal number of conjugate points $\mu_\gamma$. Thus $\nu_\gamma$ can be odd or even which results in a sine- or cosine-Fourier transform respectively. For even $k\nu_\gamma = 2m$ one has a cosine-Fourier transform which yields for (18)

$$\mathcal{F}_\gamma\{h(x)\} = \frac{(-1)^m}{4} \frac{1}{\sqrt{\pi t}} \left[\exp\left(-\frac{(x-L)^2}{4t}\right) + \exp\left(-\frac{(x+L)^2}{4t}\right)\right] \;, \tag{19}$$

whereas for odd $k\nu_\gamma = 2m+1$ the result of the sine-Fourier transform of (18) is given by

$$\mathcal{F}_\gamma\{h(x)\} = \frac{(-1)^m}{4} \frac{1}{\pi t} \left[(x-L)\,{}_1F_1\left(1, \frac{3}{2}, -\frac{(x-L)^2}{4t}\right) + (x+L)\,{}_1F_1\left(1, \frac{3}{2}, -\frac{(x+L)^2}{4t}\right)\right] \;, \tag{20}$$

where ${}_1F_1(a,b,z)$ is Kummer's function.

The Fourier transforms (19) and (20) clearly display the different kind of shapes of the periodic-orbit contributions to the trace formula as a function of $L$ for fixed $t$. If $k\nu_\gamma$ is even, one obtains a Gaussian peak at $L = k l_\gamma$, whereas for odd $k\nu_\gamma$, one has a zero at the lengths of the periodic orbits because ${}_1F_1(1, \frac{3}{2}, 0) = 1$. This is similar to the observation made by Sieber et al. [41] in case of the Fourier analysis of the spectrum.

The trace of the cosine-modulated heat kernel allows one to extract the length and stability of the first orbits and even the value of $k\nu_\gamma$ modulo 4. The last property cannot be read off in case of the often considered so-called power spectrum $D(x) = \left|\int_0^{p_{\max}} dp\, e^{ipx}[d(p) - \bar{d}(p)]\right|^2$.

With the first 11000 levels and a smoothing parameter $t = 0.0001$ we obtained the graph (full line) shown in figure 11 for the odd symmetry case. For a proper comparison with the trace formula, we have to subtract from the right-hand side of (16) the contribution of the unknown part of the energy spectrum, $\{E_n; n > N = 11000\}$. This contribution is estimated by first replacing the sum over the discrete spectrum by a Riemann-Stieltjes integral, and then approximating the integral by means of Weyl's law, equations (3) or (4):

$$\sum_{n=N+1}^\infty h(p_n) = \int_{E_N}^\infty dN(E)\, h(\sqrt{E}) \approx \int_{\sqrt{E_N}}^\infty dp\, h(p) \bar{d}(p) \;.$$

Subtracting this term from the right-hand side of (16) implies that the integration in the integral in (16) has to be restricted to the interval $[0, \sqrt{E_N}]$. With this approximation we also show in figure 11 the right-hand side of (16) (dashed line) by using the geometrical data of the primitive periodic orbits known so far (see table 3, and figure 12). We observe a striking agreement between both graphs which indicates the accuracy of the computed energy levels. Notice the excellent resolution of the two peaks near $L \approx 6.6$ which correspond to orbits No 4 and 5.

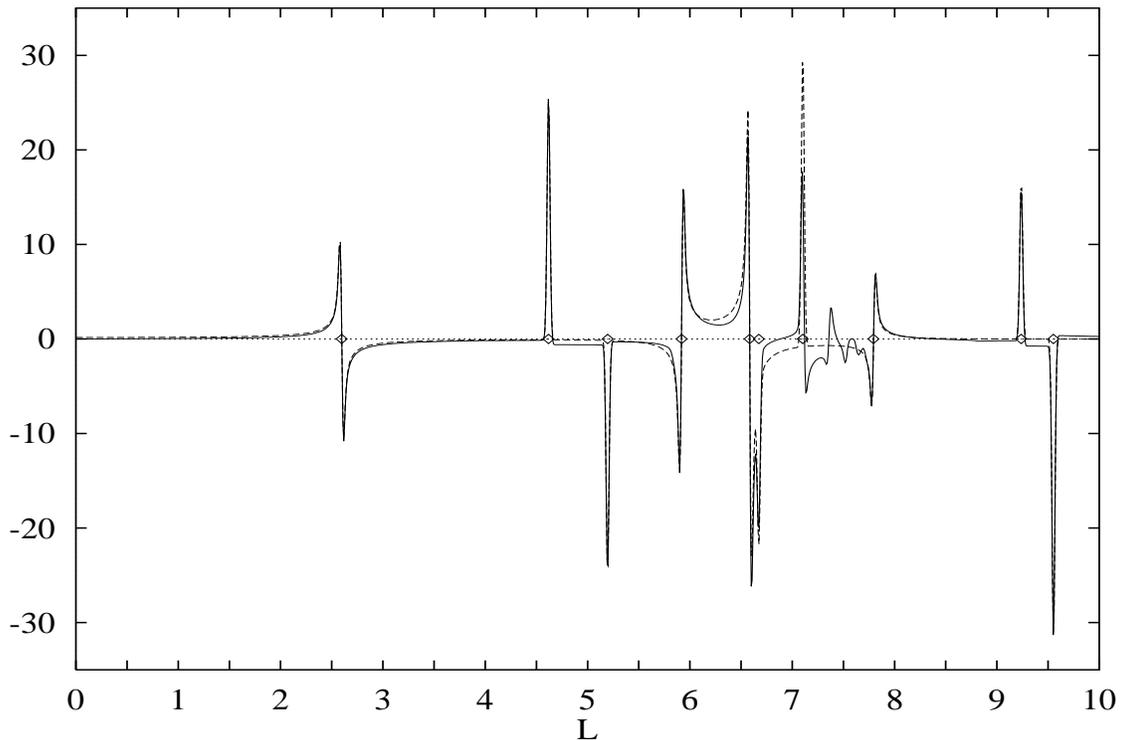

Figure 11: Trace of the cosine–modulated heat kernel for odd symmetry with $t = 0.0001$. Full line: sum over the first 11000 energy levels, dashed line: periodic-orbit sum. The squares indicate the lengths $l_\gamma$ of the first primitive periodic orbits, see table 3 and figure 12, and of their repetitions.

In the interval $L \in [7.1, 7.6]$ there are some differences between the trace of the cosine-modulated heat kernel calculated from the eigenvalues and the periodic orbit sum respectively, which are probably caused by orbits which run into the cusp of the cardioid billiard. Their contribution is expected to be of higher order in $\hbar$ and thus is not included in the trace formula (16). Similar differences exist in the even symmetry case. A detailed analysis of the trace formula for the cardioid billiard is in progress and will be presented in a separate publication.

| Orbit No | $l_\gamma$ | Tr $M_\gamma$ | $\mu_\gamma$ | $\nu_\gamma$ (D) | $\nu_\gamma$ (N) |
|---|---|---|---|---|---|
| 1 | 2.598 | -2.50 | 1 | 5 | 3 |
| 2 | 4.618 | -4.40 | 2 | 8 | 6 |
| 3 | 5.918 | -9.74 | 3 | 11 | 9 |
| 4 | 6.585 | 7.62 | 3 | 13 | 9 |
| 5 | 6.673 | -28.36 | 4 | 14 | 12 |
| 6 | 7.103 | 12.45 | 4 | 16 | 12 |
| 7 | 9.552 | -16.31 | 4 | 18 | 12 |

Table 3: Dynamical data of the first primitive periodic orbits. $l_\gamma$ is the length of the orbit $\gamma$, Tr $M_\gamma$ denotes the trace of the monodromy matrix, $\mu_\gamma$ is the maximal number of conjugate points, and $\nu_\gamma$ is the Maslov index (for odd (D) and even (N) symmetry).

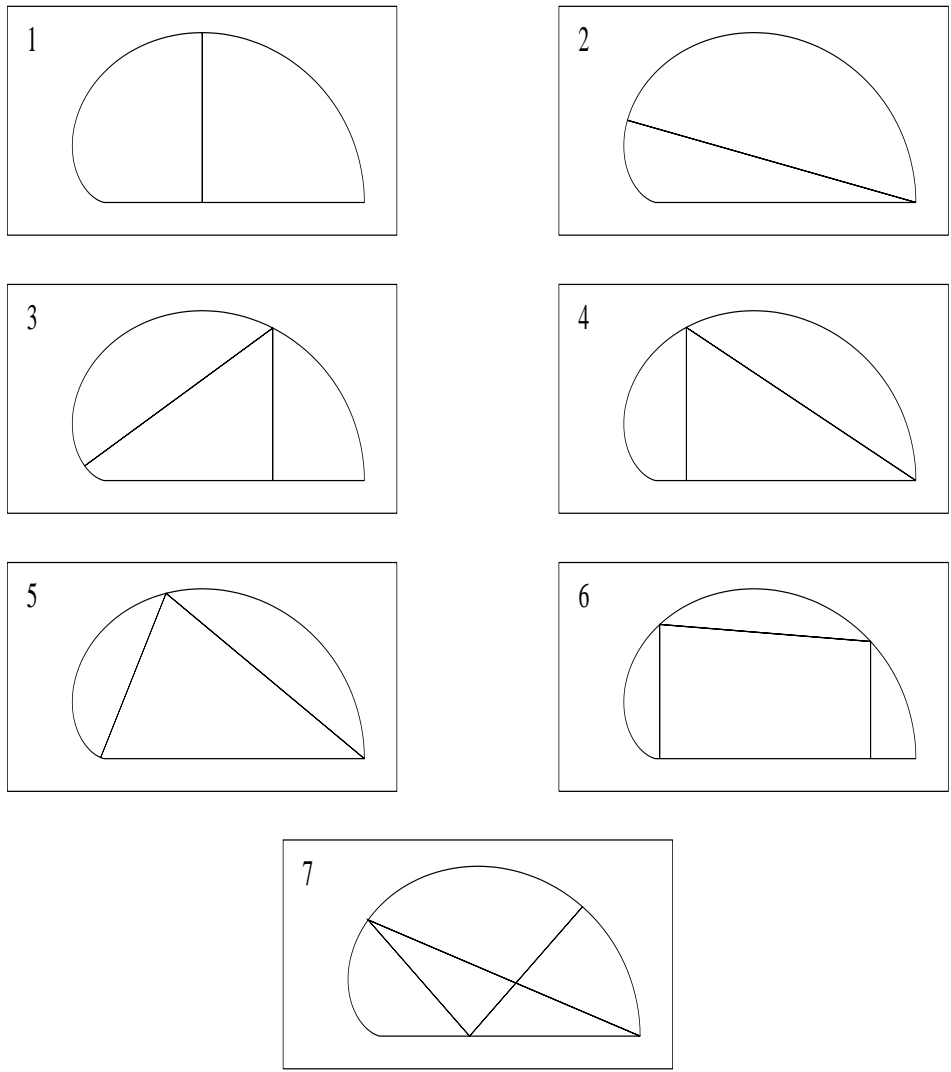

Figure 12: Primitive periodic orbits with $l_\gamma < 10$. Notice that orbit No 5 does not run into the cusp.

# 5 Summary and Discussion

In this paper we have presented a detailed analysis of the spectral statistics in the cardioid billiard. This dynamical system is strongly chaotic and thus constitutes an ideal testing ground for quantum chaology. It is found that the level-spacing distribution is in very good agreement with the Wigner distribution. However the number variance and rigidity clearly violate the universal laws of random-matrix theory for medium and large correlation lengths. The observed non-universal behaviour agrees with semiclassical expectations. Our results thus further strengthen the evidence accumulated so far showing that the applicability of the random-matrix model is limited to short-range correlations only.

A promising candidate for a universal measure of quantum chaos in spectra is the mode-fluctuation distribution $P(W)$. We have shown in sec. 3.6 that this novel spectral statistic indeed displays at a high significance level a universal Gaussian behaviour in agreement with a recent conjecture. In view of this result it appears even more urgent to inquire for a rigorous derivation of the conjecture from first principles.

Finally, in section 4, we have studied the trace of the cosine-modulated heat kernel and have compared it with the theoretical result derived from Gutzwiller's semiclassical trace formula. The overall agreement between "experiment" and theory is striking. It is amazing to see in figure 11 for instance the pronounced zero at the location of the length of the shortest periodic orbit. The observed structure follows precisely the theoretical curve as described by equation (20). This is a nice illustration of "inverse quantum chaology": Knowing the quantal eigenvalues we are able to determine not only the lengths of classical periodic orbits, but also whether they possess conjugate points or not. This agreement with periodic-orbit theory calls for a more elaborate investigation of the trace formula in the case of the quantized cardioid billiard. Work along this line is in progress and will be published later.

**Acknowledgments**

We are grateful to Marko Robnik for the kind provision of the eigenvalues of the cardioid billiard. A.B. and P.S. would like to thank Ralf Aurich and Martin Sieber for many useful discussions. Furthermore we would like to thank Jens Bolte for drawing our attention to reference [24]. One of us (P.S.) would like to thank W.P. Schleich and the University of Ulm for support.